%% file: main.tex
\documentclass[sigconf, nonacm]{acmart}

\newcommand\vldbdoi{10.14778/3421424.3421432}
\newcommand\vldbpages{61 - 73}
\newcommand\vldbvolume{14}
\newcommand\vldbissue{1}
\newcommand\vldbyear{2021}
\newcommand\vldbauthors{\authors}
\newcommand\vldbtitle{\shorttitle}
\newcommand\vldbavailabilityurl{https://github.com/neurocard}
\newcommand\vldbpagestyle{empty}

\input{macros}

\begin{document}

\newcommand{\sys}{NeuroCard\xspace}

\title{\sys: One Cardinality Estimator for All Tables}

\author{Zongheng Yang}
\affiliation{%
  \institution{UC Berkeley}
}
\email{zongheng@berkeley.edu}
\author{Amog Kamsetty}
\authornote{Equal contribution.}
\affiliation{%
  \institution{UC Berkeley}
}
\email{amogkamsetty@berkeley.edu}
\author{Sifei Luan}
\authornotemark[1]
\affiliation{%
  \institution{UC Berkeley}
}
\email{lsf@berkeley.edu}
\author{Eric Liang}
\affiliation{%
  \institution{UC Berkeley}
}
\email{ericliang@berkeley.edu}
\author{Yan Duan}
\author{Xi Chen}
\affiliation{%
  \institution{Covariant}
}
\email{{rocky, peter}@covariant.ai}
\author{Ion Stoica}
\affiliation{%
  \institution{UC Berkeley}
}
\email{istoica@berkeley.edu}

\begin{abstract}

\input{abstract}
\end{abstract}

\maketitle

\pagestyle{\vldbpagestyle}
\begingroup\small\noindent\raggedright\textbf{PVLDB Reference Format:}\\
\vldbauthors. \vldbtitle. PVLDB, \vldbvolume(\vldbissue): \vldbpages, \vldbyear.\\
\href{https://doi.org/\vldbdoi}{doi:\vldbdoi}
\endgroup
\begingroup
\renewcommand\thefootnote{}\footnote{\noindent
This work is licensed under the Creative Commons BY-NC-ND 4.0 International License. Visit \url{https://creativecommons.org/licenses/by-nc-nd/4.0/} to view a copy of this license. For any use beyond those covered by this license, obtain permission by emailing \href{mailto:info@vldb.org}{info@vldb.org}. Copyright is held by the owner/author(s). Publication rights licensed to the VLDB Endowment. \\
\raggedright Proceedings of the VLDB Endowment, Vol. \vldbvolume, No. \vldbissue\ %
ISSN 2150-8097. \\
\href{https://doi.org/\vldbdoi}{doi:\vldbdoi} \\
}\addtocounter{footnote}{-1}\endgroup

\ifdefempty{\vldbavailabilityurl}{}{
\vspace{.3cm}
\begingroup\small\noindent\raggedright\textbf{PVLDB Artifact Availability:}\\
The source code, data, and/or other artifacts have been made available at \url{\vldbavailabilityurl}.
\endgroup
}

\input{intro}
\input{approach}

\input{construction}
\input{sampling}
\input{factorization}

\input{querying}

\input{eval}
\input{related}
\input{conclusion}

\begin{acks}
   We thank Joe Hellerstein for fruitful discussions and guidance, and Michael Whittaker, Richard Liaw, and Chenggang Wu for their insightful comments on this paper.
\end{acks}

\bibliographystyle{ACM-Reference-Format}
\bibliography{refs}

\end{document}

%% file: macros.tex
\usepackage{graphicx}
\usepackage{balance}  %
\pdfoutput=1

\usepackage{enumitem}
\usepackage{framed}
\usepackage{cprotect}
\usepackage{amstext}
\usepackage{pdfpages}
\usepackage{alltt}
\usepackage{epstopdf}
\usepackage{xspace,colortbl}
\usepackage{multicol}
\usepackage{algorithm}
\usepackage{algorithmicx}
\usepackage[noend]{algpseudocode}

\usepackage[font={small}]{caption}

\usepackage{bbold}
\usepackage{dsfont}
\usepackage{bm}
\usepackage{csquotes}
\usepackage{courier}

\usepackage{float}
\usepackage{listings}
\newfloat{lstfloat}{htbp}{lop}
\floatname{lstfloat}{Listing}
\definecolor{dkgreen}{rgb}{0,0.6,0}
\definecolor{ltgray}{rgb}{0.5,0.5,0.5}

\usepackage{adjustbox}

\definecolor{light-gray}{gray}{0.95}
\definecolor{mid-gray}{gray}{0.85}
\definecolor{green}{RGB}{0,176,80}
\definecolor{darkred}{rgb}{0.7,0.25,0.25}
\definecolor{darkgreen}{rgb}{0.15,0.55,0.15}
\definecolor{darkblue}{rgb}{0.1,0.1,0.5}
\definecolor{orange}{RGB}{237,125,49}
\definecolor{blue}{RGB}{68,114,196}
\definecolor{pop}{RGB}{0,21,245}
\definecolor{myblue}{RGB}{59,106,144}
\definecolor{realblue}{RGB}{0,0,255}
\definecolor{airforceblue}{rgb}{0.36, 0.54, 0.66}
\definecolor{LightCyan}{rgb}{0.88,1,1}

\definecolor{newblue}{RGB}{162,212,236}

\usepackage{xcolor}
\usepackage{soul}
\sethlcolor{LightCyan}

\usepackage{tikz}
\usepackage{verbatim}
\usepackage{subcaption}
\newsavebox{\imagebox}
\usepackage{booktabs}
\usepackage{tabularx}

\DeclareCaptionType{Example}

\usepackage{pifont}%

\usepackage{bbm}

\newcommand{\secref}[1]{\S\ref{#1}}

\newcommand{\jobl}{\textsf{JOB-light}\xspace}
\newcommand{\joblr}{\textsf{JOB-light-ranges}\xspace}
\newcommand{\jobm}{\textsf{JOB-M}\xspace}

\let\oldsim\sim
\renewcommand{\sim}{{\oldsim}}

\newcommand{\etal}{{\emph{et al.}\,}}

\newcommand{\removedfromrevision}[1]{}
\newcommand{\removedfromcamready}[1]{}
{\endlist}

\def\ojoin{\setbox0=\hbox{$\bowtie$}%
  \rule[-.02ex]{.25em}{.4pt}\llap{\rule[\ht0]{.25em}{.4pt}}}

\def\fullouterjoin{\mathbin{\ojoin\mkern-5.8mu\bowtie\mkern-5.8mu\ojoin}}

\makeatletter
\newcommand{\distas}[1]{\mathbin{\overset{#1}{\kern\z@\sim}}}%
\newsavebox{\mybox}\newsavebox{\mysim}
\newcommand{\distras}[1]{%
  \savebox{\mybox}{\hbox{\kern3pt$\scriptstyle#1$\kern3pt}}%
  \savebox{\mysim}{\hbox{$\sim$}}%
  \mathbin{\overset{#1}{\kern\z@\resizebox{\wd\mybox}{\ht\mysim}{$\sim$}}}%
}
\makeatother

%% file: abstract.tex
Query optimizers rely on accurate cardinality estimates to produce good execution plans.
Despite decades of research, existing cardinality estimators are inaccurate for complex queries, due to making lossy modeling assumptions and not capturing inter-table correlations.
In this work, we show that it is possible to learn the correlations across all tables in a database without any independence assumptions.
We present \sys, a join cardinality estimator that builds a single neural density estimator over an entire database.
Leveraging join sampling and modern deep autoregressive models, \sys makes no inter-table or inter-column independence assumptions in its probabilistic modeling.
\sys achieves orders of magnitude higher accuracy than the best prior methods (a new state-of-the-art result of $8.5\times$ maximum error on \jobl), scales to dozens of tables, while being compact in space (several MBs) and efficient to construct or update (seconds to minutes).

%% file: intro.tex
\section{Introduction}
\label{sec:intro}

\begin{figure}[t]
\centering
\includegraphics[width=\columnwidth]{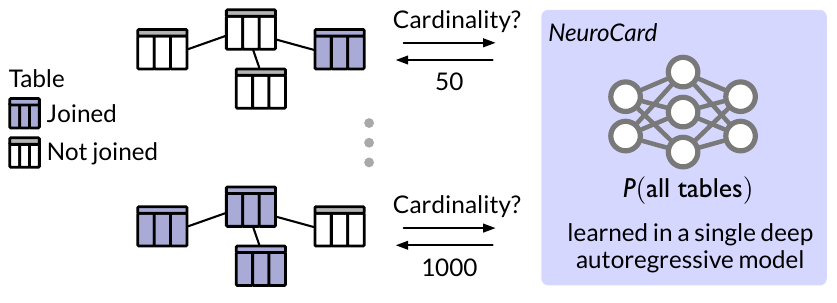}
\vspace{-.3in}
\caption{\sys uses a single probabilistic model, which learns all possible correlations among all tables in a database, to estimate join queries on any subset of tables.
  \label{fig:teaser}}
\vspace{-.2in}
\end{figure}

Query optimizers translate queries into executable plans with the best estimated performance.
They are critical not only for relational databases, but also for modern analytics engines, such as Spark \cite{sparksql} and Presto \cite{presto}.
Among various techniques, \emph{cardinality estimation} often plays a larger role than the cost model or the plan search space in producing high-quality query plans \cite{leis2015good}.
Unfortunately, cardinality estimation is a notoriously difficult problem, where the accuracy may drop exponentially as the query complexity (e.g., the number of joins) increases~\cite{leis2018query}.

At a high level, there are two approaches to cardinality estimation: \emph{query-driven} and \emph{data-driven}.  Query-driven estimators
typically rely on supervised learning to learn a function mapping (featurized) queries to predicted cardinalities.
They implicitly assume queries from a production workload are ``similar'' to training queries---namely, training and test sets of queries are drawn from the same underlying distribution.
This assumption can be violated when, for example, users issue unexpected types of queries. %

In contrast, data-driven estimators approximate the data distribution of a table---a function mapping each tuple to its probability of occurrence in the table---instead of training on ``representative'' queries. A simple method to approximate the data distribution is a histogram. In theory, once we estimate the distribution of each table in a schema, we can estimate the output cardinality of any query. While this approach is more general, it suffers from two  drawbacks: (1) \emph{lossy modeling assumptions} (e.g., assume the tables' distributions are independent), and (2) \emph{low precision} (e.g., a limited number of histogram bins). Fortunately, recent advances in machine learning have alleviated both drawbacks. Unlike previous density estimators, \emph{deep autoregressive (AR) models}~\cite{pixelcnnpp,wavenet,gpt2,made,resmade} can learn complex high-dimensional data distributions without independence assumptions, achieving state-of-the-art results in both precision and expressiveness.
This has resulted in new data-driven cardinality estimators based on deep AR models~\cite{naru}.

However, despite their promise, deep autoregressive model-based cardinality estimators %
are limited to handling single tables. There are three challenges that make this approach ineffective for \emph{joins}:
\begin{itemize}[leftmargin=*]
\item \textbf{\emph{High training cost:}} %
To learn the distribution of a join, any data-driven estimator needs to see actual tuples from the join result. Unfortunately, for all but the smallest scale, it is expensive, and sometimes infeasible, to precompute the join.
\item \textbf{\emph{Lack of generality:}}
  The AR approach builds a probabilistic model for each join, e.g., $T = T_1 \bowtie T_2 \bowtie T_3$, that it estimates.
  However, the model for $T$ cannot be directly used to estimate a join on a subset of $T$, e.g., $T_2 \bowtie \sigma(T_3)$.
  Of course, one could train a model for every possible join.
  This can be prohibitive, as the number of possible joins is exponential in the number of tables.
\item \textbf{\emph{Large model size:}}
  The complexity of the learned AR model grows with the cardinality of the dataset.
  As joins tend to involve columns with high cardinalities, an AR model built on a join may incur a prohibitively large size.%
\end{itemize}

In this work, we propose \sys, a learning-based join cardinality estimator that directly learns from data to overcome these challenges.
\sys's distinctive feature is the ability to capture the correlations across
multiple joins %
in a single deep AR model,  without any independence assumptions (Figure~\ref{fig:teaser}).  %
Once trained, this model can handle all queries issuable to the schema, regardless of what subset of tables is involved.
We address the above challenges using the following key ingredients.

To reduce training cost,
\sys samples from a join, instead of computing the join fully
(\secref{sec:sampling}). The key property of such a sample is to capture the join's distribution: if a key is more frequent in the join result, it should be more frequent in the sample as well.
To meet this requirement, we precompute the correct sampling weights for each key.
While the worst-case cost of computing the join is exponential in the number of tables,
computing the sampling weights is done in time linear with the data size by dynamic programming.

To achieve generality, \sys needs to train a single model to answer queries on any subset of tables (\secref{sec:querying}).
The basic idea behind our solution is to train the AR model on samples from the \emph{full outer join} of all tables. The full join contains the values from all the base tables, so it has sufficient information to answer a query touching any subset of tables. At inference time, if a table in the schema is not present in a join query, we need to account for any potential \emph{fanout} effect.
Consider an AR model trained on samples from the full join $T = T_1 \fullouterjoin T_2$, and a query $\sigma(T_1)$ whose cardinality we want to estimate. If the join key of $T_2$ is the foreign key of $T_1$, then a tuple of $T_1$ may \emph{appear multiple times} in $T$. \sys learns the probabilities of these ``duplicated'' tuples and additional bookkeeping information, which enables us to account for fanouts. %

Finally, to scale to large-cardinality columns while avoiding prohibitively large models, \sys employs \emph{lossless column factorization} (\secref{sec:factorization}).
An AR model stores one embedding vector per distinct value, so it could quickly blow up in size for columns with large numbers of distinct values, e.g., 100,000s or more. With factorization, a column is decomposed into several subcolumns, each taking a chunk of bits from the binary representation of the original column values.
For instance, a 32-bit ID column \textsf{id} can be decomposed into $(\textsf{id}_0, \dots, \textsf{id}_3)$ with the first subcolumn corresponding to the first 8 bits, and so on. We then train the autoregressive model on these lower-cardinality subcolumns instead of the full columns.

By combining these ingredients, \sys achieves state-of-the-art estimation accuracy, including in the challenging tail quantiles. On the popular \jobl benchmark, a schema that contains 6 tables and basic filters, \sys achieves a maximum Q-error of $8.5 \times$ using $4\,\text{MB}$. This corresponds to a $4.6\times$ improvement over the previous state of the art.
We created a more difficult benchmark, \joblr, with a larger variety of content columns and range filters.
On this benchmark, \sys achieves up to
15--34$\times$
higher accuracy than previous solutions, including DeepDB~\cite{deepdb}, MSCN~\cite{kipf2018learned}, and IBJS~\cite{ibjs_cite}.
Lastly, to test \sys's ability to handle a more complex join schema, we created \jobm which has 16 tables and multi-key joins. \sys scales well to this benchmark, offering $10\times$ higher accuracy than conventional approaches while maintaining a low model size ($27\,\text{MB}$, covering 16 tables).

\begin{figure}[t]
\centering
\includegraphics[width=.7\columnwidth]{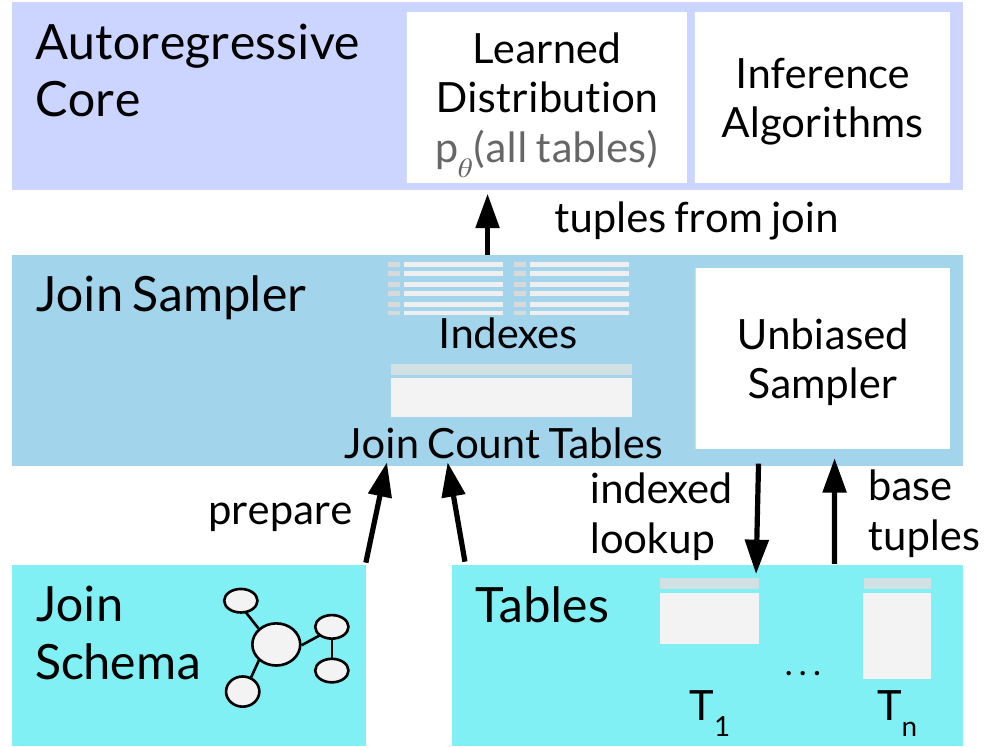}
\caption{Overview of \sys.  The Join Sampler (\secref{sec:sampling}) provides correct training data (sampled tuples from join) by using unbiased join counts. Sampled tuples are streamed to an autoregressive model for maximum likelihood training (\secref{sec:construction}).  Inference algorithms (\secref{sec:querying}) use the learned distribution to estimate query cardinalities.\label{fig:overview}}
\vspace{-.1in}
\end{figure}

In summary, this paper makes the following contributions:
\begin{itemize}[leftmargin=*]
\item We design and implement \sys, the first learned data-driven cardinality estimator that learns across joins {in a schema} without any independence assumptions. All {in-schema} correlations among the tables are captured by a single autoregressive model, which can estimate any query on any subset of tables.

  \item \sys learns the correct distribution of a join without actually computing that join. Instead, the model is trained on uniform and independent samples of the join of all tables in a schema.

  \item We propose lossless column factorization (\secref{sec:factorization}), a technique that significantly reduces the size of the autoregressive model, making its use practical for high-cardinality columns. %

  \item Compared to the best prior methods, \sys significantly improves the state-of-the-art accuracy on the \jobl benchmark. We further propose two new benchmarks, \joblr and \jobm, and show that both are much more challenging and thus better gauges of estimator quality (\secref{sec:eval}).

\end{itemize}
{To invite further research, \sys and the benchmarks used in this paper are open source at \url{https://github.com/neurocard}.}

%% file: approach.tex
\section{Overview of \sys}
\label{sec:overview}
Consider a set of tables, $T_1, \dots, T_N$.  We define their \emph{join schema} as the graph of join relationships, where vertices are tables, and each edge connects two joinable tables. %
A query is a subgraph of the overall schema. If a query joins a table multiple times, our framework duplicates that table in the schema.
We assume the schema and queries submitted to the estimator are acyclic (\secref{sec:sampling-comparison} discusses relaxations), so they can be viewed as trees.

Next, we present an overview of \sys as a sequence of goals and solutions to achieve these goals.

\subsection{Goals and Solutions}
\label{sec:requirement-solution}
\noindent {\bf Goal: A single estimator}.  Our goal is building a single cardinality estimator for the entire join schema.
For example, assuming the schema has three tables, the estimator should handle joins on any subset of tables, e.g.,  $\sigma(T_2), T_1 \bowtie T_3$, or $T_1 \bowtie T_2 \bowtie \sigma(T_3)$.

Having a single estimator has two key benefits: simplicity and accuracy. Having multiple estimators---each covering a specific join template (a table subset)---does not scale for a large number of tables, as the number of possible join templates increases exponentially.
In addition, it is easier for a DBMS to operationalize a single estimator rather than many estimators.
Most importantly, having multiple estimators can hurt accuracy.
This is because estimating the cardinality of a query on a table subset not covered by any single estimator, but by multiple estimators, requires some form of independence assumption to combine these estimators. If the independence assumption does not hold, the accuracy will suffer.

\textbf{\emph{Solution:}} We build a single cardinality estimator that learns the distribution of \emph{the full outer join of all tables} in the schema (henceforth, full join). For example, for a three-table schema, we learn $p(T_1 \fullouterjoin T_2 \fullouterjoin T_3)$. Note that using the inner join instead of the full join would not work. Indeed, the inner join $T_1 \bowtie T_2 \bowtie T_3$
is the intersection of the three tables. If a query uses only $T_1$ or $T_1 \bowtie T_3$, their tuples may not be fully contained in this intersection, and thus the estimator would have insufficient information to answer these queries.

\vspace{6pt} \noindent {\bf Goal: Efficient sampling of the full join}.
A data-driven estimator learns a distribution by reading representative tuples from that distribution.
To learn the distribution of the full join, a straightforward approach is to compute it and then uniformly draw random samples from the result.
Unfortunately, even on a small 6-table schema (the \jobl workload), the full join contains \emph{two trillion} ($2 \cdot 10^{12}$) tuples, making it infeasible to compute in practice.

\textbf{\emph{Solution:}}
We perform uniform sampling over the full join \emph{without} materializing it.
Specifically, we ensure that any tuple in the full join $J$ (a multiset) is sampled with same probability, $1/|J|$.  %
To achieve this, we leverage a state-of-the-art join sampling algorithm~\cite{zhao_sampling} (\secref{sec:sampling}).
We first precompute \emph{join count tables} that map each table's join keys to their correct sampling weights with respect to the full join.
Then, we sample the keys using these counts as weights.
Given a sampled key, we construct the full tuple by looking up the remaining columns via indexes\footnote{Like prior work on join sampling~\cite{ibjs_cite,WanderJoin}, we assume base tables have an index built for each join key. This impacts the efficiency but not correctness of the design.} from all tables, and then concatenating them.
This way, we only need to materialize the join counts as opposed to the full join. %
Using dynamic programming, computing the join counts takes time linear in the size of the database, and is quite fast in practice (e.g., 13 seconds for 6 tables in \jobl, and 4 minutes for 16 tables in \jobm). %

\vspace{6pt} \noindent {\bf Goal: Support any subset of tables}.
Although the full outer join  contains all information of the tables, we need to take care when a query involves just a subset of the tables.  Consider: %
\begin{gather*}
  T_1.\text{id}: [1, 2] \quad
  T_2.\text{id}: [1, 1] ~~ \longrightarrow ~~
  T_1 \fullouterjoin T_2: [(1,1), (1,1), (2,\varnothing)] \\
  \text{Query:}\quad \sigma_{\text{id}=1}(T_1)
\end{gather*}
The correct selectivity is $\frac{1}{2}$ (1 row).
However, in the full join distribution, $P(T_1.\text{id} = 1) = \frac{2}{3}$ (2 rows). This is because we have not accounted for the \emph{fanout} produced by the missing table, $T_2$.

\textbf{\emph{Solution:}} Handle \emph{schema subsetting}:
If a query does not include a table, we downscale the estimate by the fanout introduced by that table.
In essence, since the learned probability space is the full join, we must downscale appropriately when a query touches a subset and expects the returned selectivity to refer to that subset.

\vspace{6pt} \noindent {\bf Goal: Accurate density estimation}.
The final ingredient to achieve our goal is an accurate and compact density estimator.

\textbf{\emph{Solution:}} We leverage \emph{deep autoregressive (AR) models} to implement our density estimator.
This family of neural density estimators have been successfully employed on high-dimensional data types such as image~\cite{pixelcnnpp}, audio~\cite{wavenet}, and text~\cite{gpt2}.
Recently, Naru~\cite{naru} has leveraged deep AR models to achieve state-of-the-art accuracy results on estimating the cardinalities of single-table queries,
while \emph{learning the correlations among all columns} without independence assumptions.
We apply Naru to learn the distribution of the full join, and optimize its construction and inference for our setting.

\subsection{Putting It All Together}
\label{sec:all-together}
Figure~\ref{fig:overview} shows the high-level architecture of \sys.

Building the estimator consists of two stages. First, we prepare the join sampler by building or loading existing single-table indexes on join keys and computing the join count tables for the specified join schema (\secref{sec:sampling}).
Second, we train the deep AR model by repeatedly requesting batches of sampled tuples from the sampler, usually 2K tuples at a time.
The sampler fulfills this request in the background, potentially using multiple sampling threads.

Once the estimator is built, it is ready to compute the cardinality estimates for given queries. For each query, we use probabilistic inference algorithms (\secref{sec:querying}) to compute the cardinality estimate by (1) performing Monte Carlo integration on the learned AR model,
and (2) handling schema subsetting.
A single estimator can handle queries joining any subset of tables, with arbitrary range selections.

%% file: construction.tex
\section{Constructing \sys}
\label{sec:construction}
In this section, we present the background of the techniques used to implement \sys.

\subsection{Probabilistic Modeling of Tables}
Consider a table $T$ with column domains $\{A_1, \dots, A_n \}$.
This table induces a discrete \emph{joint data distribution}, defined as the probability of occurrence of each tuple ($f(\cdot)$ denotes number of occurrences):
\[
p(a_1, \dots, a_n) = f(a_1, \dots, a_n) / |T|.
\]
The $n$-dimensional data distribution (the \emph{joint}) $p(\cdot)$ allows us to compute a query's cardinality  as follows.
Define a query $Q$ as $\sigma: A_1 \times \cdots \times A_n \rightarrow \{0, 1\}$.
Then, the \emph{selectivity}---the fraction of records that satisfy the query---can be computed as a probability: %
  $P(Q) = \sum_{a_1 \in A_1} \cdots \sum_{a_n \in A_n} \sigma(a_1, \dots, a_n) \cdot p(a_1, \dots, a_n)$.
  The \emph{cardinality} is obtained by multiplying it with the row count: $|Q| = P(Q) \cdot |T|$.

Data-driven cardinality estimators can be grouped along two axes: (1) joint factorization, and (2) the density estimator used.

{\bf Joint factorization}, or the modeling assumption, determines how precisely data distribution $p$ is factored. %
Any modeling assumption
risks
losing information about correlations across columns, which ultimately leads to a loss in accuracy.
For example, the widely used %
ID histogram technique assumes the columns are independent. As a result, it factors $p$ into a set 1D marginals, $p \approx \prod_{i=1}^{n}p(A_i)$, which can lead to large inaccuracies when the columns' values are strongly correlated.
Similarly, other data-driven cardinality estimators such as graphical models~\cite{getoor2001selectivity,getoor2001learning,deshpande2001independence,tzoumas2013efficiently,tzoumas2011lightweight} either assume conditional independence or partial independence among columns. One exception is the autoregressive (product-rule)   factorization, %
\begin{gather}
p = \prod_{i=1}^{n} p(A_i | A_{<i}),
\end{gather}
which precisely expresses the overall joint distribution as the product of the $n$ conditional distributions.

{\bf The density estimator} determines how precisely the aforementioned factors
are actually approximated.
The most accurate ``estimator'' would be recording these factors exactly in a hash table. Unfortunately, this leads to enormous construction and inference costs (e.g., in the case of $p(A_n | A_{1:n-1})$).
At the other end, the 1D histogram has low costs, but this comes at the expense of low precision, as it makes no distinction between the values falling in the same bin. Over the years, a plethora of solutions have been proposed, including kernel density estimators and Bayesian networks.
Recently, \emph{deep autoregressive (AR) models}~\cite{resmade,gpt2,pixelcnnpp} have emerged as the density estimator of choice. Deep AR models compute $\{ p(A_i | A_{<i})\}$ without explicitly materializing them by learning the $n$ conditional distributions in compact neural networks.
Deep AR models achieve state-of-the-art precision, and, for the first time, provide a tractable solution for implementing the autoregressive factorization. %

\subsection{Naru: Deep Autoregressive Models as Cardinality Estimators}
\label{sec:naru}

\sys builds on Naru, a state-of-the-art cardinality estimator that fully captures the correlations among all columns of a single table using a deep AR model. Next, we present an overview of Naru and discuss how \sys leverages it.

\vspace{6pt} \noindent \textbf{Construction.}
Given table $T$, an AR model $\theta$ takes a tuple $\bm{x} \in T$ as input, and predicts conditional probability distributions, $\{p_\theta(X_i | \bm{x}_{<i})\}$, each of which is an 1D distribution over the $i$-th column (conditioned on all prior column values of $\bm{x}$).
The likelihood of the input tuple is then predicted as %
$p_\theta(\bm{x}) = \prod_{i=1}^n p_\theta(X_i = \bm{x}_i | \bm{x}_{<i})$.
Any deep AR architecture can instantiate this framework, e.g., ResMADE~\cite{resmade} or Transformer~\cite{transformer}.
Training aims to approximate the data distribution $p$ using $p_\theta$, by minimizing the KL divergence~\cite{murphy2012machine}, $D_{KL}(p || p_\theta)$. This is achieved by maximum likelihood estimation (MLE) and gradient ascent to maximize the predicted (log-)likelihood of data:
\begin{gather}
  \text{Sample i.i.d.} \quad \bm{x}\, \sim\, p  \label{eq:sample-iid}\\
  \text{Take gradient steps to maximize} \quad \log p_\theta(\bm{x})
\end{gather}

In our setting, we define $T$ as the full outer join of all tables within a schema. Consequently, the deep AR model learns the correlations across all tables. Next, we need to sample tuples with probabilities prescribed by $p$. Otherwise, $p_\theta$ would approximate an incorrect, biased distribution. To achieve this, we use a sampler that emits \emph{simple random samples} from the full join $T$ (\secref{sec:sampling}).

\vspace{6pt} \noindent \textbf{Estimating query cardinalities.} Once constructed, the Naru estimator estimates the cardinality of a given query.
A query is represented as a hyper-rectangle: each column $X_i$ with domain $A_i$ is constrained to take on values in a valid region $R_i \subseteq A_i$:
\begin{gather}
  \text{Query:} \quad \land \{X_i \in R_i \}
\end{gather}
Next, Naru estimates the probability of the query (an event) using a Monte Carlo integration algorithm, \emph{progressive sampling}:
\begin{gather}
  \textsf{ProgressiveSampling($\{X_i \in R_i\}$):} \quad p_\theta (\land \{X_i \in R_i \}) \cdot |T| \label{eq:est-card}
\end{gather}
It works by drawing imaginary, in-region tuples from the model's learned distributions. Specifically, it draws the first dimension of the sample as $x_1 \sim p_\theta(X_1 |X_1 \in R_1)$, the second dimension of the sample as $x_2 \sim p_\theta(X_2 |X_2\in R_2; x_1)$, and so on. The likelihoods of the samples are importance-weighted. This procedure also efficiently supports omitted columns, i.e., wildcards of the form $X_i \in \ast$.

\sys's inference invokes progressive sampling to estimate cardinalities, but extends it in two ways. First, we apply the column factorization optimization (\secref{sec:factorization}), which potentially changes a $X_{i+i}$'s valid region, $R_{i + 1}$, based on the value drawn from $X_i$. Second, we add support for schema subsetting (\secref{sec:querying}), by downscaling selectivity $p_\theta (\land \{X_i \in R_i \})$ by the corresponding fanout.

\subsection{Join Problem Formulation}
\label{sec:formulation}
A join schema induces the full outer join of all tables in the schema, $T = T_1 \fullouterjoin \cdots \fullouterjoin T_N$. Our goal is to build a fully autoregressive probabilistic model on the full join consisting of all tables' columns:
\begin{equation}
  \text{Model:} \quad p_\theta(T) \equiv p_\theta(T_1.\text{col}_1, T_1.\text{col}_2, \dots, T_N.\text{col}_k)\label{eq:prob-model}
\end{equation}
We can then use the probabilistic model to estimate the cardinalities of join queries on any subset of tables in the schema.

{
\vspace{4pt}
\noindent \textbf{Supported joins.}
\sys supports acyclic join schemas and queries containing multi-way, multi-key equi-joins (\secref{sec:sampling-comparison} discusses how to relax the acyclic requirement).
The schema should capture the most common joins.
For joins not captured in the schema,
their cardinalities can be estimated by first obtaining single-table estimates using \sys, then combining the estimates using classical heuristics~\cite{leis2018query}. This allows uncommon cases to be handled under the same framework, albeit at the cost of lower accuracy.
}

\vspace{4pt}
\noindent \textbf{Supported filters.}
\sys supports equality and range filters on discrete or numerical columns. These include arithmetic comparison operators ($<, >, \leq, \geq, =$) and \textsf{IN}.  More complex filters can also be expressed using the valid region encoding, mentioned in the previous section.
Arbitrary forms of \textsf{AND}/\textsf{OR} can be handled via the inclusion-exclusion principle.

\begin{figure}[t]
\centering
\includegraphics[width=\columnwidth]{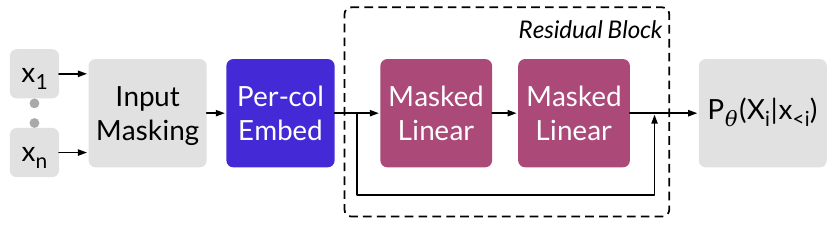}
\vspace{-.3in}
\caption{Default architecture of the autoregressive model.\label{fig:arch}}
\vspace{-.1in}
\end{figure}

\subsection{Model architecture}
\label{sec:arch}
\sys uses a standard AR architecture, ResMADE~\cite{resmade}, which is also employed by Naru; see Figure~\ref{fig:arch}.
Input tuples are represented as discrete, dictionary-encoded IDs, $(x_1, \dots, x_n)$, and embedded by per-column embedding matrices.
The concatenated embedded vector is fed to a series of residual blocks, each consisting of two masked linear layers (they are masked to ensure the autoregressive property). The output layer produces logits $\{\log p_\theta(X_i | \bm{x}_{<i}) \}$ by dotting the last layer's output with the embedding matrices. Next, we compute a cross-entropy loss on the logits and perform backpropagation. We turn on Naru's \emph{wildcard skipping} optimization, which randomly masks inputs to train special marginalization tokens that aid infer-time estimation (i.e., using these tokens to skip sampling any wildcards in a query).

Masked multi-layer perceptrons such as ResMADE strike a good balance between efficiency and accuracy.
\sys can use any advanced AR architectures, if desired.
{In \secref{sec:eval}, we also instantiate \sys with an advanced architecture (the Transformer~\cite{transformer}).}

%% file: sampling.tex
\section{Sampling from Joins}
\label{sec:sampling}

\input{fig-example2}

A key challenge in \sys is computing an \emph{unbiased} sample of the full join (\secref{sec:requirement-solution}) to ensure that the learned distribution faithfully approximates the full join distribution.
Namely, every tuple in the full join $J$ (a multiset) must be sampled equally likely with probability $1/|J|$.  The samples should also be i.i.d., as required by Equation~\ref{eq:sample-iid}.
\sys meets these requirements by using a sampler that produces \emph{simple random samples with replacement}.

\subsection{Algorithm}
\label{sec:sampling-algo}
A tuple in the full join contains \emph{join key columns} and \emph{content columns}.
Our sampler exploits this decomposition.
The first step of the sampler is to precompute \emph{join count tables}, which are per-table statistics that reflect the occurrence counts of the join keys in the full join.
The sampler then samples the join keys, table-by-table, with occurrence probabilities proportional to their join counts.
Lastly, it selects content columns from the base tables by looking up the drawn join keys.
This completes a batch of sample, which is sent to the model for training, and the procedure repeats on demand.

\vspace{4pt}
\noindent {\textbf{Computing join counts.}}
Zhao \etal~\cite{zhao_sampling} provide an efficient algorithmic framework of join sampling that produces simple random samples from general multi-key joins.
\sys implements the Exact Weight algorithm from Zhao \etal, adapted to full outer joins.

We illustrate the algorithm on a join schema (a tree) consisting of tables $T_1,\dots,T_N$.
For exposition, assume they only involve join keys (content columns are gathered later). %
Let $T_1$ be the root table.
The join count of a tuple $t \in T_i$ is the total number of tuples in the full outer join of all of $T_i$'s descendants that joins with $t$.
It is recursively defined as:
\begin{equation}
\begin{aligned}
  w_i(t) = \prod_{T_j \in \mathrm{Children}(T_i)}~~ \sum_{t' \in t \rtimes T_j} w_j(t') \quad \forall i, \forall t \in T_i
  \label{eq:join-counts}
\end{aligned}
\end{equation}
where $t \rtimes T_j$ denotes all tuples in $T_j$ that join with $t$. For a leaf table with no descendants,
$w_i(\cdot)$ is defined as 1.
At the root table $T_1$, $w_1(t)$ represents the count of all $t \in T_1$ in the entire full outer join.
{The join counts of each table are computed by aggregating over the join counts of all of its child tables, and can thus be computed recursively in a bottom-up fashion.}
Using dynamic programming, the time complexity is linear in the number of tuples in all tables, $O(|T_1| + \cdots + |T_N|)$.

\vspace{3pt}
\noindent \textbf{Sampling.}
Once the join counts are computed, the sampler produces a sample by traversing the join tree in a top-down fashion. It starts by drawing a sample $t_1$ from the root table $T_1$ using weights $\{w_1(t): t \in T_1\}$ (i.e., with probabilities $\{ w_1(t) / \sum_{t' \in T_1} w_1(t')\}$). It then samples through all descendants of $T_1$ in the breadth-first order. At a child table, say $T_2$, it samples $t_2$ from $t_1 \rtimes T_2$ (all tuples in $T_2$ that join with $t_1$) using weights $\{w_2(t): t \in t_1 \rtimes T_2\}$. The procedure continues recursively until all tables are visited, and thus produces a sample $(t_1, \cdots, t_N)$, each $t_i$ being a tuple of join keys from the respective table.

\vspace{3pt}
\noindent \textbf{Example.} Consider the schema in Figure~\ref{fig:e2e-schema}. %
Figure~\ref{fig:e2e-jct} shows the computed join counts.
The leaf table $C$ has a count of 1 for every tuple. In $B$, since $(2,c)$ can join with two tuples in $C$, its join count is $2=1+1$. Similar propagation happens for $A.x=2$ which gets a count of $3=1+2$.
Physically, we store the join counts indexed by join keys (e.g., for $C$, only one mapping $c\rightarrow 1$ is kept).
For sampling, suppose $A.x=2$ is first sampled. It has two matches in $B$ with weights 1 and 2, so the second match, $(2,c)$, has an inclusion probability of $2/3$.

\vspace{3pt}
\noindent {\bf \texttt{NULL} handling.}
To support full outer joins, we handle \texttt{NULL} keys as follows.
We add a virtual $\varnothing$ tuple (which denotes \texttt{NULL}) to each table $T_i$, and make it join with all normal $t \in T_j$ that have no matches in $T_i$,
where $T_j \in \mathrm{Children}(T_i)$. %
Similarly, any normal $t\in \mathrm{Parent}(T_i)$ that has no match in $T_i$ joins with $T_i$'s $\varnothing$. All-\texttt{NULL} is invalid. Propagation proceeds as before;
Figure~\ref{fig:e2e-jct} shows examples.

\vspace{4pt} \noindent {\bf Constructing complete sample tuples.}
In the prior example, suppose $\langle 2; 2,c; c \rangle$ is drawn.
We gather the content columns of $A$ by looking up $A.x = 2$ and similarly for $(B.x, B.y) = (2, c)$\footnote{Either intersect two matching lists from both columns' index lookups, or do a single lookup if a composite index is available.} and $C.y=c$. On multiple matches, we pick a row uniformly at random.
Their concatenation represents a sampled tuple from the full join.

\vspace{4pt} \noindent {\bf Computing the size of the full join (normalizing constant).}
Recall from \secref{sec:naru} that the row count $|J|$ (the \emph{normalizing constant} in probabilistic terms) is required to convert selectivities into cardinalities.
With join counts it can be computed exactly: $|J| = \sum_{t \in T_1} w_1(t)$.

\vspace{4pt} \noindent {\bf Parallel sampling.}
Finally, the sampling procedure is embarrassingly parallel: after the join count tables $\{w_i(\cdot)\}$ are produced, parallel threads can be launched to
read the join counts and produce samples. Computation of the join count tables is also parallelizable, although it is an one-time effort.
Sampling correctness is preserved even in the presence of parallelism due to the i.i.d. property.

\subsection{Comparison with other samplers}
\label{sec:sampling-comparison}
Our key requirements of uniform and i.i.d. samples from the full join render many related sampling algorithms unsuitable.  If either property is not satisfied, the sampling distribution would be biased and thus compromise the quality of the learned AR model.
As examples, Index-based Join Sampling (IBJS)~\cite{ibjs_cite} is neither uniform nor independent; Wander Join~\cite{WanderJoin} produces independent but non-uniform samples.
Both approaches do produce unbiased estimators for counts or other aggregate statistics, but are not designed to return uniform join samples.
Reservoir sampling, a well-known technique, draws samples without replacement (thus, non-independent) and requires a full scan over the full join, which is not scalable.
Lastly, the Exact Weight algorithm \sys implements is among the most efficient in Zhao \etal~\cite{zhao_sampling}. They provide additional extensions to support general, potentially cyclic joins (e.g., a cycle can be \emph{broken}), which \sys can leverage to broaden our formulation (\secref{sec:formulation}).

%% file: fig-example2.tex
\def\schemaGraph{\tikz[circle,inner sep=1pt,  label=below:{\small Test}] {
                  \node[draw] (A) at (0,0.8) {$A$};
                  \node[draw] (B) at (1,0) {$B$};
                  \node[draw] (C) at (2,0.8) {$C$};
                  \node (label) at (.35,.25) {\\\\\scriptsize{$x$}};
                  \node (label2) at (1.75,.25) {\\\\\scriptsize{$y$}};
                  \draw (A)--(B)--(C);
                }}%
\def\schemaGraphFlat{\tikz[circle,inner sep=1.2pt,label=above] {
                  \node[draw] (A) at (0,0) {$A$};
                  \node[draw] (B) at (1,0) {$B$};
                  \node[draw] (C) at (2,0) {$C$};
                  \draw (A)--(B) node[midway, above] {\small{$x$}};
                  \draw (B)--(C) node[midway, above] {\small{$y$}};
                }}%

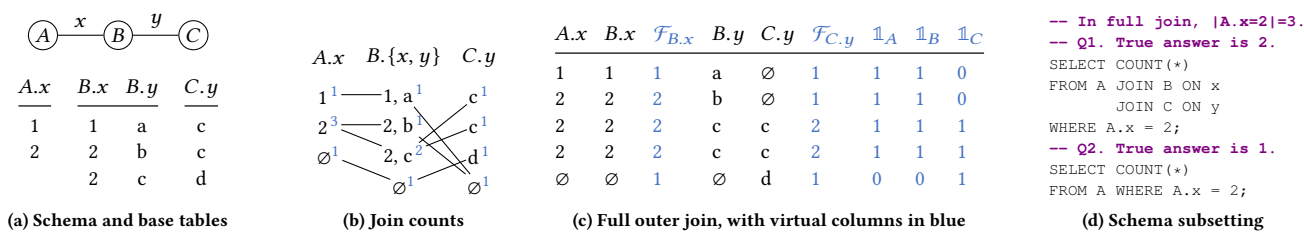
\begin{figure*}[htp]
\captionsetup[subfigure]{justification=centering}
    \centering
\begin{subfigure}[b]{.45\columnwidth}
\small
\centering
        \begin{tikzpicture}[inner sep=1pt, text centered,]
          \node at (-13,.9){
              \schemaGraphFlat
          };
          \node at (-13,-0.5) {
              \begin{tabular}{@{} c @{}}
                 $A.x$ \\ \midrule
                1 \\
                2 \\
                \\
              \end{tabular}
              \quad
\tabcolsep=0.11cm
              \begin{tabular}{@{} c c @{}}
                $B.x$ & $B.y$ \\ \midrule
                1 & a \\
                2 & b \\
                2 & c
              \end{tabular}
              \quad
            \begin{tabular}{@{} c @{}}
                  $C.y$ \\ \midrule
                c  \\
                c \\
                d \\
              \end{tabular}
          };
        \end{tikzpicture}
\caption{Schema and base tables\label{fig:e2e-schema}}
    \vspace{.2cm}
\end{subfigure}\hfill
\begin{subfigure}[b]{.3\columnwidth}
\small
\centering
        \begin{tikzpicture}[inner sep=.5pt]
        \node (A1) at (0,1.7) {$A.x$};             \node (B1) at (1,1.7) {$B.\{x,y\}$};        \node (C1) at (2,1.7) {$C.y$};

        \node (A1) at (0,1.2) {1$^{\,\textcolor{blue}{1}}$};             \node (B1) at (1,1.2) {1, a$^{\,\textcolor{blue}{1}}$};        \node (C1) at (2,1.2) {c$^{\,\textcolor{blue}{1}}$};
        \node (A2) at (0,0.8) {2$^{\,\textcolor{blue}{3}}$};               \node (B2) at (1,0.8) {2, b$^{\,\textcolor{blue}{1}}$};          \node (C2) at (2,0.8) {c$^{\,\textcolor{blue}{1}}$};
        \node (A3) at (0,0.4) {$\varnothing$$^{\textcolor{blue}{1}}$}; \node (B3) at (1,0.4) {2, c$^{\,\textcolor{blue}{2}}$};        \node (C3) at (2,0.4) {d$^{\,\textcolor{blue}{1}}$};
                                               \node (B4) at (1,0) {$\varnothing$$^{\textcolor{blue}{1}}$}; \node (C4) at (2,0) {$\varnothing$$^{\textcolor{blue}{1}}$};

                  \draw [-] (A1)--(B1) node[midway, above] {\small{}};
                  \draw [-] (B3)--(C1) node[midway, above] {\small{}};

                  \draw [-] (A2)--(B2) node[midway, above] {\small{}};
                  \draw [-] (B3)--(C2) node[midway, above] {\small{}};

                  \draw [-] (A2)--(B3) node[midway, above] {\small{}};
                  \draw [-] (B4)--(C3) node[midway, above] {\small{}};

                  \draw [-] (A3)--(B4) node[midway, above] {\small{}};
                  \draw [-] (B1)--(C4) node[midway, above] {\small{}};
                  \draw [-] (B2)--(C4) node[midway, above] {\small{}};

        \end{tikzpicture}
\caption{Join counts\label{fig:e2e-jct}}
    \vspace{.2cm}
\end{subfigure}\hfill
\begin{subfigure}[b]{.7\columnwidth}
\small
\centering
        \begin{tikzpicture}[inner sep=1pt, text centered,]

          \node at (-13,-0.5) {

\tabcolsep=0.11cm
\begin{tabular}{@{} l l l l l l lll @{}}
    $A.x$ & $B.x$ & $\textcolor{blue}{\mathcal{F}_{B.x}}$ &  $B.y$  & $C.y$  & $\textcolor{blue}{\mathcal{F}_{C.y}}$ & $\textcolor{blue}{\mathbb{1}_A}$ & $\textcolor{blue}{\mathbb{1}_B}$ & $\textcolor{blue}{\mathbb{1}_C}$\\ \midrule
  1 & 1   & \textcolor{blue}{1} & $\text{a}$  & $\varnothing_{}$ & $\textcolor{blue}{1}$ & \textcolor{blue}{1} & \textcolor{blue}{1} & \textcolor{blue}{0} \\
  2 & $2$ & \textcolor{blue}{2} & $\text{b}$  & $\varnothing$ & \textcolor{blue}{1} & \textcolor{blue}{1} & \textcolor{blue}{1} & \textcolor{blue}{0} \\
  2 & $2$ & \textcolor{blue}{2} & $\text{c}$  & $\text{c}$ & \textcolor{blue}{2}  & \textcolor{blue}{1} & \textcolor{blue}{1} & \textcolor{blue}{1} \\
  2 & $2$ & \textcolor{blue}{2} & $\text{c}$  & $\text{c}$ & \textcolor{blue}{2}  & \textcolor{blue}{1} & \textcolor{blue}{1} & \textcolor{blue}{1} \\
  $\varnothing$ & $\varnothing$ & \textcolor{blue}{1}& $\varnothing$  & $\text{d}$ & \textcolor{blue}{1} & \textcolor{blue}{0} & \textcolor{blue}{0} & \textcolor{blue}{1}
\end{tabular}

          };
        \end{tikzpicture}
\caption{Full outer join, with virtual columns in blue\label{fig:e2e-full-join}}
    \vspace{.2cm}
\end{subfigure}\hfill
\begin{subfigure}[b]{.42\columnwidth}
\small
\centering

\begin{lstlisting}[
language=SQL,
showspaces=false,
basicstyle=\ttfamily\scriptsize,
commentstyle=\color{violet}\bfseries\scriptsize,
keywordstyle=\ttfamily\scriptsize,
]
 -- In full join, |A.x=2|=3.
 -- Q1. True answer is 2.
 SELECT COUNT(*)
 FROM A JOIN B ON x
        JOIN C ON y
 WHERE A.x = 2;
 -- Q2. True answer is 1.
 SELECT COUNT(*)
 FROM A WHERE A.x = 2;
\end{lstlisting}%
    \vspace{-.2cm}
\caption{Schema subsetting\label{fig:e2e-queries}}
    \vspace{.2cm}
\end{subfigure}
        \vspace{-.3cm}
        \caption{End-to-end example.
          (a) A join schema of three tables and their join key columns. Content columns are omitted.
          (b) Join counts (blue) enable uniform sampling of the full outer join and are computed in linear time by dynamic programming. Here, edges connect join partners.
          (c) Learning target: the full outer join of the schema, with \emph{virtual columns} in blue. We show the \emph{fanouts} $\mathcal{F}$, the number of times a join key value appears in the corresponding base table, for keys $B.x$ and $C.y$.  The fanouts for $A.x$ and $B.y$ are all $1$ and omitted.  Each \emph{indicator} $\mathbb{1}_T$ denotes whether a tuple has a match in table $T$.
          \emph(d) Examples of schema subsetting, i.e., queries that touch a subset of the full join (\secref{sec:querying}).
        }
\end{figure*}

%% file: factorization.tex
\section{Lossless Column Factorization}
\label{sec:factorization}
A key challenge of using an autoregressive model for high-cardinality data is that the size of the model parameters can scale linearly with the numbers of distinct values in the columns.
In the model architecture we use (\secref{sec:arch}), each column (any data type; categorical or numerical) is first dictionary-encoded into integer token IDs.  Then a per-column \emph{embedding} layer is applied on these token IDs.
The size of the trainable embedding matrix (essentially, a hash table) for each column $C$ scales linearly with $|C|$, i.e., the number of distinct values in the domain.
Even a moderately sized column with up to $10^6$ distinct values, therefore, easily takes up $128\,\text{MB}$ of space, assuming 32-dimensional embeddings are used.

\begin{figure}[t]
\centering
\includegraphics[width=\columnwidth]{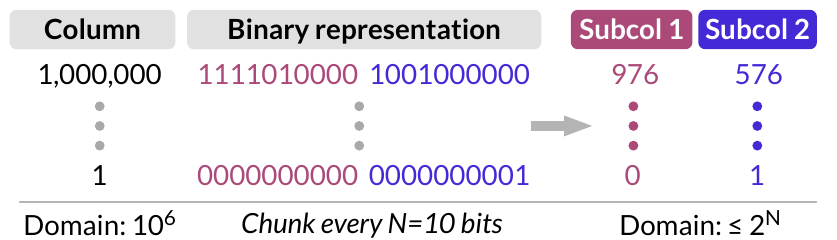}
\vspace{-.25in}
\caption{Lossless column factorization (\secref{sec:factorization}).\label{fig:factorization}}
\vspace{-.2in}
\end{figure}

To handle high-cardinality columns efficiently, we propose an optimization that we call \emph{lossless column factorization}.
This optimization is inspired by the popular use of ``subword units''~\cite{nlp_subword} in modern natural language processing, and also shares characteristics with ``bit slicing'' in the indexing literature~\cite{bit_slicing}.
Different from subword units, column factorization does not use a statistical algorithm such as byte pair encoding to determine what subwords to use (a potential optimization).  Different from bit slicing, we slice a value into groups of bits and convert them back into base-10 integers.

Figure~\ref{fig:factorization} illustrates the idea on a simple example.
Suppose a column (any datatype) has a domain size of $|C| =10^6$.  Naively supporting this column would require allocating $|C| \cdot h$ floats as its embedding matrix, where $h$ is the embedding dimension.
Instead, \sys factorizes each value \emph{on-the-fly} during training: we convert an original-space value into its binary representation, then slice off every $N$ bits, the \emph{factorization bits} hyperparameter.   Each sliced off portion becomes a \emph{subcolumn}, now in base-10 integer representation.  These subcolumns are now treated as regular columns to learn over by the autoregressive model.
 Crucially, a much smaller embedding matrix is now needed for each subcolumn containing at most $2^N \cdot h$ floats. In this example, we can reduce $128\,\text{MB}$ to $250\,\text{KB}$---a more than $500\times$ space reduction.

\vspace{6pt}
\noindent {\bf Model size vs. statistical efficiency.}
Choosing the factorization bits $N$ enables a tradeoff between model size vs. statistical efficiency.
By decreasing $N$, we have more subcolumns, each with a smaller domain, but learning across more variables becomes harder. %
In theory, by using autoregressive modeling no information is lost in this translation, so the precision of the learned distributions is not affected.
In practice, we observed that lower factorization bits, i.e., slicing into more subcolumns, generally underperform higher ones that use more space, but not by a significant margin (\secref{sec:ablation}).
We thus set the factorization bits $N$ based on a space usage budget.

 \vspace{6pt} \noindent {\bf Lossless = factorization + autoregressive modeling.}  With factorization, a column is factorized into multiple subcolumns, which are then fed into a downstream density estimator.  However, if a density estimator with independence assumptions, e.g., 1D histograms, is used, then this whole process is \emph{lossy}.  By modeling $p(\textsf{subcol}_1, \textsf{subcol}_2) \approx p(\textsf{subcol}_1) p(\textsf{subcol}_2)$, histograms would fail to capture any potential correlation between the two subcolumns.
In other words, other estimators \emph{could} read in subcolumn values and potentially reduce space usage, but their inherent quality and assumptions determine how much information is learned about the subcolumns, and about their correlations with other columns.
 By using autoregressive modeling, \sys forces the AR model to explicitly capture such correlation, namely (ignoring other columns):
\[
  p(\textsf{col}) \equiv p(\textsf{subcol}_1, \textsf{subcol}_2) = p(\textsf{subcol}_1)p(\textsf{subcol}_2 | \textsf{subcol}_1),
\]
which has no inherent loss of information.
Hence, we  call the unique combination of factorization and autoregressive modeling \emph{lossless}.

 \vspace{6pt}
 \noindent {\bf Filters on subcolumns.} During probabilistic inference, a filter on an original column needs to be translated into \emph{equivalent} filters on subcolumns.
 Recall from \secref{sec:naru} that the probabilistic inference procedure draws samples that lie inside the queried region.
 We modify that procedure to handle subcolumns by respecting each filter's semantics.
Going back to our example, consider the filter $\textsf{col} < 1{,}000{,}000$.
The filter for the high-bits $\textsf{subcol}_1$ is \emph{relaxed} to $\leq 976$ (note the less-equal).
The inference procedure would draw a $\textsf{subcol}_1$ value in this range, based on which the low-bits filter is relaxed appropriately.
If the drawn $\textsf{subcol}_1$ is $976$, then the filter on $\textsf{subcol}_2$ is set to  ``$< 576$''; otherwise, the high-bits already satisfy the original filter so a wildcard is placed on the low-bits subcolumn.  This is reminiscent of processing range predicates on bit-sliced indexes~\cite{bit_slicing}; \sys applies these processing logic in the  new context of probabilistic inference for autoregressive models.

%% file: querying.tex
\section{Querying \sys}
\label{sec:querying}

Once built, the autoregressive model summarizes the entire full outer join.
The challenge with querying this probabilistic model for a selectivity estimate is that the query may \emph{restrict the space it touches to a subset of the full join}---a phenomenon we term \emph{schema subsetting}.
Since the selectivity estimate returned by the model assumes the probability space to be the full outer join, rather than the query-specific restricted space, the estimate should be \emph{downscaled} appropriately during probabilistic inference.

\sys's inference algorithms combine two building blocks.
First, Naru~\cite{naru} introduced \emph{progressive sampling}, a Monte Carlo algorithm that integrates over an autoregressive model to produce selectivity estimates.
We invoke this routine (i.e., Equation~\ref{eq:est-card}) on the trained autoregressive model with changes outlined in this section.
Second, Hilprecht \etal~\cite{deepdb} have proposed inference algorithms to query a sum-product network trained on a full outer join.
We state their algorithms below and discuss how to adapt these algorithms into our framework, thereby generalizing them to a new type of probabilistic model.

\vspace{5pt} \noindent {\bf Basic case: no table omitted.}
The simplest case of schema subsetting is an \emph{inner} join query on all tables.
Consider the example data in Figure~\ref{fig:e2e-schema} and an inner join query Q1 in Figure~\ref{fig:e2e-queries}.
The query, $\sigma_{A.x=2}(A \bowtie_x B \bowtie_y C)$, restricts the probability space from the full join to the inner join.
Naively querying the model for $|A.x=2|$ would return a cardinality of $|J| \cdot (3/5) = 3$ rows, as 3 out of 5 rows in the full join $J$ (Figure~\ref{fig:e2e-full-join}) satisfy the filter.
However, the correct row count for this query is $2$ (two rows in the inner join; both pass the filter).
Left/right outer joins can also exhibit this behavior.

To correct for this, Hilprecht \etal~propose a simple solution by adding an \emph{indicator column} per table into the full join.  A binary column $\mathbb{1}_T$ is added for each table $T$, with value $1$ if a tuple (in the full join) has a non-trivial join partner with table $T$, and $0$ otherwise.

\sys adopts this solution as follows. %
First, during training, the sampler is tasked with appending these \emph{virtual} indicator columns on-the-fly to sampled tuples.
Recall that each sampled tuple is formed by querying base-table indexes with sampled join keys.
If a table $T$ contains a join key, we set that sampled tuple's $\mathbb{1}_T$ to $1$, and $0$ otherwise (see Figure~\ref{fig:e2e-full-join}).
The autoregressive model treats these indicator columns as regular columns to be learned.

Second, during inference, \sys adds equality constraints on the indicator columns, based on what tables are present in the query.  The progressive sampling routine (Equation~\ref{eq:est-card}) not only gets the usual filter conditions, $\{X_i \in R_i\}$, but also $\{ \mathbb{1}_T = 1\}$ for any table $T$ that appears in the inner-join query graph\footnote{The indicator columns can also be constrained appropriately for left or right joins.}.  In summary, for the no-omission case, the routine now estimates the probability:%
\begin{gather}
  P(\{X_i \in R_i\} \land \{ \mathbb{1}_T = 1: \text{for all table $T$}\})
\end{gather}

\noindent {\bf Example.}  Coming back to the example query Q1, $\sigma_{A.x=2}(A \bowtie_x B \bowtie_y C)$, we compute the selectivity under the full join as
$P(A.x = 2 \land \mathbb{1}_A = \mathbb{1}_B = \mathbb{1}_C = 1)$.
Reading from Figure~\ref{fig:e2e-full-join}, this probability is $2/5$, so the cardinality is correctly computed as $5 \cdot (2/5) = 2$ rows.

\vspace{5pt} \noindent {\bf Omitting tables and fanout scaling.}
The less straightforward case is if a query \emph{omits}, i.e., does not join, certain tables.  Consider Q2 in Figure~\ref{fig:e2e-queries}: $\sigma_{A.x=2}(A)$.  When restricting the scope to table $A$, the row count of $A.x=2$ is 1, different from $|J| \cdot P(A.x = 2 \land \mathbb{1}_A=1) = 3$ rows.
The fundamental reason this happens is because the operation of a full join has \emph{fanned out} tuples from base tables.  To correctly downscale, Hilprecht \etal~propose recording a per-join \emph{fanout} column.  We adapt this solution in \sys\footnote{Our definition differs slightly from  Hilprecht \etal. In that work, each fanout column is bound to a PK-FK join and stores the frequency of a value in the FK. Our treatment binds a fanout to each join key, regardless of PK/FK, and is defined as the frequency each value appears in that key column itself. This removes their assumption of PK-FK joins and supports general equi-joins where both join keys can have duplicate values.}.

Specifically, for each join key column $T.k$, we insert into the full join a virtual fanout column, $\mathcal{F}_{T.k}$, defined as the number of times each value appears in $T.k$.
For example, $2$ appears twice in $B.x$, so its fanout is $\mathcal{F}_{B.x}(2) = 2$; see Figures~\ref{fig:e2e-schema} and \ref{fig:e2e-full-join}.  Again, we task the join sampler with adding these fanout values on-the-fly
to each batch of sampled tuples.  The inclusion of fanouts is piggybacked onto the index lookup path (querying the size of each lookup result list), which adds negligible overheads.

On the inference side, Hilprecht \etal~showed that the correct cardinality with omitted tables can be computed via \emph{fanout scaling}:
\begin{equation}
\begin{aligned}
  \text{Cardinality(query Q)}
  &= |J| \cdot P(\{X_i \in R_i\} ~~ \text{subsetted to query Q}) \\
  &= |J| \cdot \underset{X \sim J}{\mathbb{E}} \left[ \frac{\mathbb{1}_{\{X_i \in R_i\}} \cdot \prod_{T \in Q} \mathbb{1}_T} {\prod_{R \notin Q} \mathcal{F}_{R.\text{key}}} \right].
\end{aligned}
\label{eq:fanout-scaling}
\end{equation}
In essence, the numerator handles the basic case above, while the denominator counts the total number of times omitted tables $\{R \notin Q\}$ have fanned out each tuple in query Q.
It loops through each omitted table $R$, finds its unique
join key $R.\text{key}$  that connects to Q in the schema (discussed in detail below), and looks up the associated fanout value $\mathcal{F}_{R.\text{key}}$.
We incorporate this scaling as follows.
Since the fanout columns are learned by the model, we modify progressive sampling to draw a concrete value for each relevant $\mathcal{F}_{R.\text{key}}$ per progressive sample, compute the product of these fanouts, and divide the progressive sample's estimated likelihood by this product.

\vspace{5pt}
\noindent {\bf Example.} Coming back to Q2, $\sigma_{A.x=2}(A)$, the constraints are $\{A.x = 2, \mathbb{1}_A = 1\}$.  Reading from Figure~\ref{fig:e2e-full-join}, three rows satisfy the constraints and the relevant downscaling keys are $B.x$ and $C.y$. Thus the expectation expands as: $\frac{1}{5} \cdot (\frac{1}{2\cdot1} + \frac{1}{2\cdot2}+\frac{1}{2\cdot2}) = \frac{1}{5}$. Multiplying with $|J|=5$ arrives at the correct cardinality of $1$ row.

\vspace{5pt}
\noindent {\bf Handling fanout scaling for multi-key joins.}  Our formulation of fanout scaling supports multi-key joins, e.g., both $x$ and $y$ keys in the example schema $A.x = B.x \land B.y = C.y$ (Figure~\ref{fig:e2e-schema}).  The challenge of fanout scaling in this case is determining the set of omitted keys to downscale.  %
Let $V$ be the set of all tables.
Let $Q$ be the set of tables joined in a query, and the complement $O = V \setminus Q$ the omitted tables.  Pick any table $T \in Q$.  There exists a unique path from each omitted $T_O \in O$ to $T$, because the join schema graph is a tree (acyclic, connected).  The join key attached to the edge incident to $T_O$ on this path is the unique join key for table $T_O$ to downscale.  Hence, the fanout downscaling factor in Equation~\ref{eq:fanout-scaling} is well-defined.

Going back to example Q2 where only $A$ is queried, when considering the omitted table $B$ which has two join keys ($B.x$, $B.y$), we see that $B.x$ is the unique fanout key since it lies on the path $A\longleftrightarrow B$.

\vspace{5pt} \noindent {\bf Summary of schema subsetting.}
To recap, \sys's probabilistic inference leverages the progressive sampling algorithm from Naru and the idea of additional columns from Hilprecht \etal~that we term \emph{virtual columns}.  Our join sampler is modified to logically insert into the full join two types of virtual columns, the indicators and the fanouts.  Both are treated as regular columns to be learned over by the density model, and both are used during progressive sampling to handle various cases of schema subsetting.

\vspace{5pt} \noindent {\bf Ordering virtual columns in the autoregressive factorization.}  The autoregressive model requires some fixed ordering of columns in its factorization (\secref{sec:naru}).  Naru has shown that different orderings may have different performance in the tail error but not in the lower error quantiles.  We adopt the same practice as Naru in using an arbitrary ordering for the content columns.  For the virtual columns introduced above, we place them after all the content columns, with indicators before fanouts.
The intuition here is to ensure that (1) the conditional distributions involving content columns do not get confused by the presence of virtual columns, and (2) when sampling fanouts, placing them at the end allows for prediction using a maximum amount of prior information.

In our early benchmarks this choice performed better than if virtual columns were placed early in the ordering.  We also experimented with \emph{multi-order training}~\cite{made} in the autoregressive model, but did not see noticeably better performance.  Thus, we opt for a simple treatment and leave such optimizations to future work.

%% file: eval.tex
\begin{figure}[tp]
\centering
\includegraphics[width=.95\columnwidth]{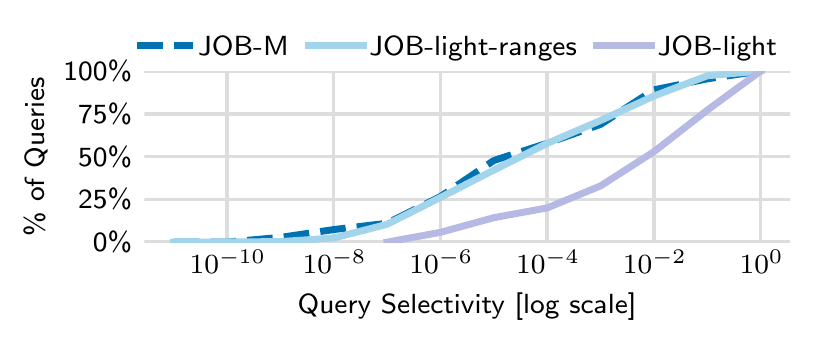}
\vspace{-.2in}
\caption{\small Distribution of query selectivity (\secref{sec:eval-workloads}).\label{fig:query-sel}}
\vspace{-.1in}
\end{figure}

\section{Evaluation}
\label{sec:eval}
We evaluate \sys on accuracy and efficiency and compare it with state-of-the-art cardinality estimators. The key takeaways are:
\begin{itemize}[leftmargin=*]
\item \textbf{\sys outperforms the best prior methods by
    4--34$\bm{\times}$
    in accuracy} (\secref{sec:accuracy}).
  On the popular \jobl benchmark, \sys achieves a maximum error of $8.5\times$ using $4\,\text{MB}$.
\item \textbf{\sys scales well to more complex queries} (\secref{sec:accuracy}). On the two new benchmarks \joblr (more difficult range filters) and \jobm (more tables in schema), \sys achieves orders of magnitude higher accuracy than prior approaches.
  \item \textbf{\sys is efficient to construct and query} (\secref{sec:efficiency}).  A few million tuples, learned in less than 5 minutes, suffice for it to reach best-in-class accuracy.
  \item \textbf{We study the relative importance of each component of \sys} (\secref{sec:ablation}).  Out of all factors, learning the correlations across all tables and performing unbiased join sampling prove the most impactful. %
\end{itemize}

\subsection{Experimental Setup}
\label{sec:eval-workloads}

\begin{table}[tp]\centering \small%
\caption{{Workloads used in evaluation. \emph{Tables}: number of base tables. \emph{Rows, Cols, Dom.}: row count, column count, and maximum column domain size of the full outer join of each schema.  \emph{Feature} characterizes each workload's queries. Rows in full join: $\bm{2\cdot10^{12}; 2\cdot10^{12}; 10^{13}}$.}\label{table:workloads}}
\vspace{-.1in}
\begin{tabular}{@{} l l l l l  l @{}} \toprule
{\textsc{Workload}} & {\textsc{Tables}}& {\textsc{Rows}} & {\sc Cols} & {\sc Dom. }   & \textsc{Feature}  \\ \midrule
\jobl & 6 & $2\cdot10^{12}$ & 8 &  235K & single-key joins \\
\joblr & 6 & $2\cdot10^{12}$ & 13 & 134K & $+$complex filters \\
\jobm & 16 & $10^{13}$ & 16 & 2.7M & $+$multi-key joins \\
\bottomrule
\end{tabular}
\end{table}

\noindent {\bf Workloads} (Table~\ref{table:workloads}).
We adopt the real-world IMDB dataset and schema to test cardinality estimation accuracy.
Prior work~\cite{leis2018query,leis2015good} reported that correlations abound in this dataset and established it to be a good testbed for cardinality estimators.
We test the following query workloads on IMDB:
\begin{itemize}[leftmargin=*]
\item {\bf \jobl}: a 70-query benchmark used by many recent cardinality estimator proposals~\cite{e2e_cost,kipf2018learned,deepdb}.
The schema contains 6 tables, {\sf title} (primary), {\sf cast\_info}, {\sf movie\_companies}, {\sf movie\_info}, {\sf movie\_keyword}, {\sf movie\_info\_idx} and is a typical star schema---every non-primary table only joins with {\sf title} on {\sf title.id}.
The full outer join contains $2\cdot 10^{12}$ tuples.
Each query joins between 2 to 5 tables, with only equality filters except for range filters on {\sf title.production\_year}.

\item {\bf \joblr}: we synthesized this second benchmark containing 1000 queries derived from \jobl by enriching filter variety.
We generate the 1000 queries uniformly distributed to each join graph of \jobl (18 in total), as follows.
For each join graph, using our sampler we draw a tuple from the inner join result.
We use the non-null column values of this tuple as filter literals, and randomly place 3--6 comparison operators associated with these literals, based on whether each column can support range (draw one of $\{\leq, \geq, =\}$) or equality filters ($=$).
Overall, this generator (1) follows the data distribution and guarantees non-empty results, and (2) includes more filters, in variety and in quantity, than \jobl.
An example 3-table query is: $\textsf{mc} \bowtie \sigma_\textsf{\scriptsize{info\_type\_id=99}}(\textsf{mi\_idx}) \bowtie \sigma_{\textsf{\scriptsize{episode\_nr}}\leq 4 \land \textsf{\scriptsize{phonetic\_code}} \geq \textsf{\scriptsize{'N612'}}}(t)$, where \textsf{t.id} is joined with other tables' \textsf{movie\_id}.
\item {\bf \jobm}: this last benchmark contains 16 tables in IMDB and involves \emph{multiple} join keys. For instance, the table {\sf movie\_companies} is joined not only with {\sf title} on {\sf movie\_id}, but also with {\sf company\_name} on {\sf company\_id}, and with {\sf company\_type} on {\sf company\_type\_id}, etc. We adapt the 113 JOB queries~\cite{leis2015good} by allowing each table to appear at most once per query and removing logical disjunctions (e.g., {\sf A.x=1 $\vee$ B.y=1}). Each query joins 2--11 tables.
  We use \jobm to test \sys's scalability as its full join is $5\times$ larger and has more dimensions than the above (see Table~\ref{table:workloads}).

\end{itemize}
{The benchmarks are available at \url{https://github.com/neurocard}.}

\vspace{4pt}
\noindent {\bf Metric.} We report the usual Q-error distribution of each workload, where the Q-error of a query is the multiplicative factor an estimated cardinality deviates from the query's true cardinality:
$\text{Q-error(query)} := \max \left( \frac {\textsf{card}_{\textsf{actual}}}  {\textsf{card}_{\textsf{estimate}}}, \frac {\textsf{card}_{\textsf{estimate}}}  {\textsf{card}_{\textsf{actual}}} \right)$.
Both actual and estimated cardinalities are lower bounded by 1, so the minimum attainable Q-error is $1\times$.  As reported in prior work~\cite{naru}, reducing high-quantile errors is much more challenging than mean or median; thus, we report the quantiles $p100, p99, p95$, and the median.
For timing experiments, we report latency/throughput using an AWS EC2 VM with a NVIDIA V100 GPU and 32 vCPUs. %

\vspace{4pt}
\noindent {\bf Benchmark characteristics.}
Figure~\ref{fig:query-sel} plots the distributions of selectivities of these workloads, where we calculate each query's selectivity as ${\textsf{card}_{\textsf{actual}}} / {\textsf{card}_{\textsf{inner}}}$ (denominator is the row count of the query join graph---an inner join---without filters).
The selectivity spectrums of our two benchmarks (\joblr and \jobm) are much wider than \jobl due to higher filter variety.
The median selectivity is more than $100\times$ lower, while at the low tail the minimum selectivities are $1000\times$ lower.

\subsection{Compared Approaches}
\label{sec:baselines}

We compare against several prevalent families of estimators. In each family, we aim to choose a state-of-the-art representative.
Related Work (\secref{sec:related}) includes a more complete discussion on all families and their representative methods.

\vspace{4pt}
\noindent {\bf Supervised query-driven estimators.}
We use MSCN~\cite{kipf2018learned} as a recent representative from this family.
It takes in a featurized query, runs the query filters on pre-materialized samples of the base tables, then use these bitmaps as additional network inputs, and predicts a final cardinality.
For \jobl, we used the training queries and sample bitmaps provided in the authors' source code~\cite{mscn-code}.
For \joblr, due to new columns, we generated 10K new training queries---generating and executing them to obtain true cardinality labels took 3.2 hours---and used a bitmap size of 2K to match the size of other estimators in this benchmark.
For \jobl, we also cite the best numbers obtained by Sun and Li~\cite{e2e_cost}, termed \emph{E2E}, which is a deep supervised net with more effective building blocks (e.g., pooling, LSTM) than MSCN.

\vspace{4pt}
\noindent {\bf Unsupervised data-driven estimators.}
We use DeepDB~\cite{deepdb} as a recent technique in this family.
It uses a non-neural sum-product network~\cite{poon2011sum} as the density estimator for each table subset {chosen by correlation tests}. Conditional independence is assumed across subsets.
{In contrast, \sys uses a neural autoregressive model to build a single learned estimator over all tables in a schema.}
We use two recommended configurations from DeepDB: a base version that learns up to 2-table joins, and a larger version that additionally builds 3-table models.
{These correspond to their storage-optimized and the standard setups, respectively.}

We found that the DeepDB source code~\cite{deepdb-code} did not support range queries on categorical string columns out-of-the-box.  Since \joblr contains such queries, we perform data and query rewriting for this baseline, by dictionary-encoding the string values into integers.  Reported results are with this optimization enabled.

\vspace{4pt}
\noindent {\bf Join sampling.}
We implement %
the Index-based Join Sampling method (IBJS)~\cite{ibjs_cite}, using 10,000 as the maximum sample size.
A query's cardinality is estimated by taking a sample from the query's join graph and executing per-table filters on-the-fly.

\vspace{4pt}
\noindent {\bf Real DBMS.}  We use Postgres v12, which performs cardinality estimation using 1D histograms and heuristics to combine them.

\vspace{4pt}
\noindent {\bf Other baselines.} The methods chosen above have been compared to other estimators in prior studies.
Naru~\cite{naru} has shown that estimators based on classical density modeling (KDE; Bayesian networks; the MaxDiff n-dimensional histogram) or random sampling significantly lag behind deep autoregressive models.
DeepDB~\cite{deepdb} also shows that it significantly outperforms wavelets~\cite{wavelets}.
We therefore do not compare to these methods.

\vspace{4pt}
\noindent {\bf \sys.} We implement \sys on top of the Naru source code~\cite{naru_source}.
{We use ResMADE by default. For complex benchmarks we also use the Transformer (\secref{sec:arch}), which is suffixed with \textsf{-large}.}

\begin{table}[t]\centering \small
\caption{\jobl, estimation errors. Lowest errors are bolded. \label{table:jobl-error}}
\vspace{-.1in}
\begin{tabular}{@{} l l l l l l l l @{}} \toprule
\textsc{Estimator} & {\sc Size} && {Median}  & {95th} & {99th} & {Max} \\ \midrule
Postgres & \(70\,\text{KB}\) &&7.97 & 797 & $3\cdot 10^3$ & $10^3$ \\
IBJS & -- && 1.48 & $10^3$ & $10^3$ & $10^4$ \\

MSCN & \(2.7\,\text{MB}\) && 3.01 & 136 & $1 \cdot 10^3$ & $10^3$ \\

  E2E (quoting~\cite{e2e_cost}) & N/A %
                                && 3.51 & 139 & 244 & 272 \\
DeepDB & \(3.7\,\text{MB}\) && 1.32 & 4.90 & 33.7 & 72.0 \\
DeepDB-large & \(32\,\text{MB}\) && {\bf 1.19} & \textbf{4.66} & 35.0 & 39.5 \\\midrule

\sys & \(3.8\,\text{MB}\) && {1.57} & {5.91} & \textbf{8.48} &  \textbf{8.51} \\
\bottomrule
\end{tabular}
\end{table}

\subsection{Estimation Accuracy}
\label{sec:accuracy}
\subsubsection{\jobl}

Table~\ref{table:jobl-error} reports each estimator's accuracy on the 70 \jobl queries.
Overall, \sys exhibits high accuracy across the spectrum.
\textbf{It sets a new state-of-the-art maximum error at $\bm{8.5\times}$} using $3.8\,\text{MB}$ of parameters.
This represents an $>8\times$ improvement over the best prior method when controlling for size.

We now discuss a few observations.
Not surprisingly, Postgres has the most inaccurate median---indicating a systematic mismatch between the approximated distribution and data---due to its use of coarse-grained density models (histograms) and heuristics.
IBJS fares better at the median, but falls off sharply at tail,
because samples of a practical size have a small chance to hit low-density queries in a large joint space.
Both MSCN and E2E are deep supervised regressors which show marked improvements over prior methods.  However, their median and 95th errors are quite similar and have sizable gaps from the two data-driven estimators.

\sys vs. DeepDB shows interesting trends. \sys is up to 4--8$\times$ better at tail (99th, max), and DeepDB is slightly better at lower quantiles.
\sys is more robust at tail due to (1) a markedly better density model (neural autoregressive vs. non-neural sum-product networks that use inter-column independence assumptions), and (2) learning all possible correlations among the columns of all 6 tables, whereas DeepDB assumes (conditional) independence across several table subsets. DeepDB-large, being $8.4\times$ bigger and trained on $7.7\times$ more (54M) tuples, still trails \sys at tail by more than $4\times$. %
\sys slightly trails at the lower quantiles (``easy'' queries with high true density) likely due to the mode-covering behavior of KL-divergence minimization~\cite{goodfellow}.%

\subsubsection{\joblr}  This 1000-query benchmark adds equality/range filters on more content columns, using the same join templates as \jobl (which has range filters on one column only).
Results are shown in Table~\ref{table:joblr-error}.

\textbf{\sys achieves the best accuracy across all error quantiles, and improves on the best prior methods by up to $\bm{15{-}34\times}$.}
It is also the only estimator with $<2\times$ median and 3-digit 99\%-tile errors.
Overall, all estimators produce less accurate cardinalities, though the drops are of varying degrees.
Compared with MSCN, \sys improves by $2\times$ at median, $7\times$ at 95th, $15\times$ at 99th, and $2\times$ at max.
Compared with DeepDB, \sys improves the four quantiles by $2\times$, $9\times$, $21\times$, and $23\times$, respectively.
Comparing the enlarged versions of the two estimators (suffixed with \textsf{-large}),
the accuracy gains become $1.4\times, 2.6\times, 9.6\times$ and $34\times$,
respectively.

\sys's improvements over baselines significantly widen in this benchmark, due to prior approaches failing to capture the more complex inter-column correlations being tested.

\begin{table}[t]\centering \small
\caption{\joblr, estimation errors. Lowest errors bolded.\label{table:joblr-error}}
\vspace{-.1in}
\begin{tabular}{@{} l l l l l l l l @{}} \toprule
\textsc{Estimator} & {\sc Size} && {Median}  & {95th} & {99th} & {Max} \\ \midrule
Postgres & \(70\,\text{KB}\) && 13.8 & $2\cdot 10^3$ & $2 \cdot 10^4$ & $5 \cdot 10^6$ \\
IBJS & -- && 10.1 & $4\cdot 10^4$ & $10^6$ & $10^8$ \\

MSCN & \(4.5\,\text{MB}\) && 4.53  &  397 & $6\cdot 10^3$ & $2\cdot 10^4$ \\

DeepDB & \(4.4\,\text{MB}\) && 3.40 & 537 & $8 \cdot 10^3$ & $2 \cdot 10^5$ \\

  {DeepDB-large} & \(21\,\text{MB}\) && 2.00 & 91.7 & $2\cdot 10^3$ & $4 \cdot 10^4$ \\

  \midrule

\sys & \(4.1\,\text{MB}\) && {1.87} & {57.1} & {375} & {8169} \\

  {\sys-large} & \(21\,\text{MB}\) && \textbf{1.40} & \textbf{35.1} & \textbf{232} & \textbf{1029} \\

\bottomrule
\end{tabular}
\end{table}

\begin{table}[t]\centering \small
\caption{\jobm, estimation errors. Lowest errors are bolded.\label{table:jobm-error}}
\vspace{-.1in}
\begin{tabular}{@{} l l l l l l l l @{}} \toprule
\textsc{Estimator} & {\sc Size} && {Median}  & {95th} & {99th} & {Max} \\ \midrule
Postgres & \(120\,\text{KB}\) && 174 & $1\cdot 10^4$ & $8\cdot 10^4$ & $1 \cdot 10^5$ \\
IBJS & -- && 61.1 & $3\cdot 10^5$ & $4\cdot 10^6$ & $4\cdot 10^6$ \\
\midrule

  {\sys} & \(27.3\,\text{MB}\) && \textbf{2.84} & \textbf{404} & \textbf{1327} & $\bm{2\cdot 10^4}$ \\

  {\sys-large} & \(409\,\text{MB}\) && \textbf{1.96} & \textbf{26.4} & \textbf{304} & \textbf{874}  \\

\bottomrule
\end{tabular}
\end{table}

\subsubsection{\jobm}
This final benchmark tests \sys's ability to scale to a much larger and more complex join schema. Different from the \jobl schema, \jobm contains 16 tables, with each query joining 2--11 tables on multiple join keys (in addition to {\sf movie\_id} only in \jobl). For baselines, we only include Postgres and IBJS, because MSCN's query encoding does not support the complex filters in this benchmark and DeepDB ran out of memory
on this 16-table dataset due to high-cardinality categorical columns.

Results in Table~\ref{table:jobm-error} show that \textbf{\sys's accuracy remains high on this complex schema}.
Postgres produces large errors, and IBJS also struggles, due to many intermediate samples becoming empty as the number of joins grows.
\sys overcomes this challenge and offers more than $10\times$ better accuracy across the board.
In terms of space efficiency, since the model needs to be trained on the full outer join of 16 tables and the maximum domain size exceeds 2 million, a vanilla \sys would require $900\,\text{MB}$ in model size.
With column factorization (\secref{sec:factorization}), the model size is reduced to 27MB---less than 1\% of the total size of all tables.
{We also present a large model \sys-large to demonstrate scalability.}

\begin{figure}[t]
     \centering
     \begin{subfigure}[b]{0.235\textwidth}
         \centering
          \includegraphics[width=\columnwidth]{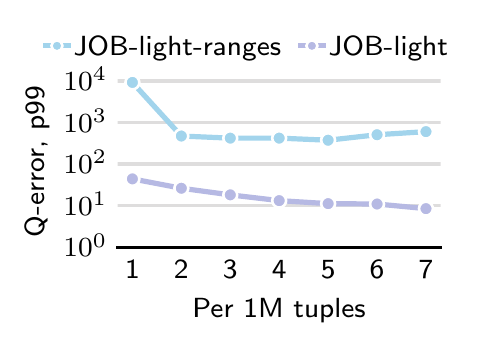}
\vspace{-.23in}
         \caption{\small{Accuracy vs. Tuples Trained}\label{fig:learning-curve}}
     \end{subfigure}
     \begin{subfigure}[b]{0.235\textwidth}
         \centering
          \includegraphics[width=\columnwidth]{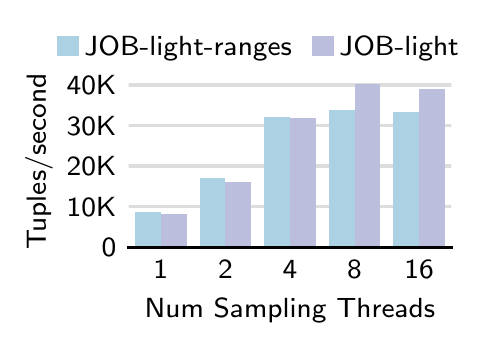}
\vspace{-.23in}
         \caption{\small{Training Throughput}\label{fig:train-throughput}}
     \end{subfigure}
     \begin{subfigure}[b]{0.237\textwidth}
         \centering
          \includegraphics[width=\columnwidth]{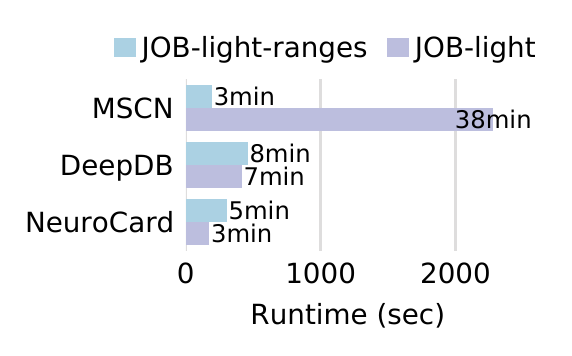}
\vspace{-.23in}
         \caption{\small{Training Comparison}\label{fig:train-time-compare}}
     \end{subfigure}
     \begin{subfigure}[b]{0.235\textwidth}
         \centering
          \includegraphics[width=\columnwidth]{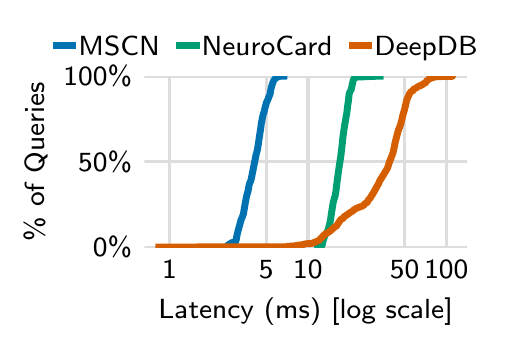}
\vspace{-.23in}
         \caption{\small{Inference Comparison}\label{fig:infer-cdf}}
     \end{subfigure}
\caption{\small {Statistical and physical efficiency of \sys.}
\vspace{-.15in}
\label{fig:runtime-routing}}
\end{figure}

\subsection{Efficiency}
\label{sec:efficiency}
Having established that \sys achieves the best accuracy, we now study the statistical and physical efficiency of \sys.

\vspace{4pt}
\noindent {\bf How many tuples are required for good accuracy?}  Figure~\ref{fig:learning-curve} plots accuracy (p99 on \jobl and \joblr) vs. number of tuples trained.  \emph{About 2--3M tuples are sufficient for \sys to achieve best-in-class accuracy} (compare with Tables~\ref{table:jobl-error} and \ref{table:joblr-error}).  Using more samples helps, but eventually yields diminishing returns.
Reaching high accuracy using a total of $\sim 10^7$ samples out of a population of $~10^{12}$ data points (i.e., only 0.001\% of the data)---many queries would inevitably touch unseen data points---shows that \sys generalizes well and is \emph{statistically efficient}.

\vspace{4pt} \noindent
{\bf How does sampling affect training throughput?} Figure~\ref{fig:train-throughput} plots the training throughput, in tuples per second, vs. the number of sampling threads used to provide training data.  Four threads suffice to saturate the GPU used for training.
At lower thread counts, the device spends more time waiting for training data than doing computation.
With a peak throughput of $\sim 40$K tuples/second, \sys can finish training on 3M tuples in about 1.25 minutes.

\vspace{4pt} \noindent
{\bf Wall-clock training time comparison.} Figure~\ref{fig:train-time-compare} compares the wall-clock time used for training the MSCN, DeepDB, and \sys configurations reported in Tables~\ref{table:jobl-error} and \ref{table:joblr-error}.
MSCN requires a separate phase of executing training queries to collect true cardinalities, which takes much longer (3.2 hours for 10K queries) than just the training time shown here.
{DeepDB runs on parallel CPUs and is quite efficient.}
{\sys starts training/on-the-fly sampling after calculating the join count tables, which takes 13 seconds for both datasets.}
Its construction is efficient due to parallel sampling and accelerated GPU computation.

\vspace{4pt} \noindent
{\bf Wall-clock inference time comparison.} Lastly, Figure~\ref{fig:infer-cdf} plots the latency CDF of the learning approaches for 1000 \joblr queries. As before, we use the base configurations reported in the accuracy Tables.
MSCN and \sys run on GPU while DeepDB runs on CPU; all three approaches are implemented in Python. %
MSCN is fastest because its lightweight network has fewer calculations involved.
DeepDB's latencies span a wide spectrum, from $\sim 1\,\text{ms}$ for queries with low complexity (numbers of joins and filters involved) to $\sim 100\,\text{ms}$ for queries with the highest complexity.
\sys's latencies are more predictable, with $17\,\text{ms}$ at median and $12\,\text{ms}$ at minimum: this is due to the higher number of floating point operations involved in the neural autoregressive model.
All approaches can be sped up by engineering efforts (e.g., if run in a native language). For \sys, model compression or weight quantization can also reduce the computational cost.

\begin{table}[t]\centering \small%
  \caption{{Ablation studies: varying primary components of \sys. Unlisted values are identical to the Base configuration.
      We show the impact of the sampler (A), column factorization bits (B), autoregressive model size (C), inter-table correlations learned (D), and whether to use an autoregressive model at all (E) on the 50\% and {95}\%-tile errors of \joblr.}\label{table:ablation}}
\vspace{-.1in}
\begin{tabular}{@{} >{\centering}m{0.06\linewidth} c >{\centering}m{0.05\linewidth} >{\centering}m{0.11\linewidth} >{\centering}m{0.23\linewidth}  cc c @{}} \toprule
  & Sampler & Fact. Bits & $d_{\text{ff}}; d_{\text{emb}}$ & Correlations Learned &&  {\bf p50} %
  & {\bf p95} \\ \midrule
  Base {\scriptsize(4.1\,MB)} & unbiased & 14& 128; 16 & all tables in one AR && 1.9 %
                                                                                     & {57.1} \\ \midrule

  (A) & biased &   &  & && 33 %
                              & {$3270$} \\ \midrule

  (B) & &  10 {\scriptsize(2.2\,MB)} &  & && 2.2 %
                                                 & {173}\\
                       & &  12 \scriptsize{(2.6\,MB)}  &  & && 2.0 %
                                                                   & {168} \\
                       & &  None {\scriptsize(12\,MB)}  &  & && 1.6 %
                                                                    & {62.7}\\ \midrule

  (C) & &  & 128; 64 {\scriptsize(23\,MB)} & && 1.5 %
                                                    & {44.0}\\

                       & &  & 1024; 16 {\scriptsize(31\,MB)} & && 1.7 %
                                                                      & {64.0}\\ \midrule

  (D)  & &  &       & one AR per table && 40 %
                                             & {$9\cdot10^4$}\\ \midrule

  (E) & \multicolumn{5}{c}{No model; uniform join samples only} & 4.0 %
                                                                      & {$2\cdot 10^5$}\\
\bottomrule
\end{tabular}
\vspace{-.1in}
\end{table}

\subsection{Dissecting \sys}
\label{sec:ablation}

To gain insights, we now evaluate the relative importance of primary components of \sys, by varying them and measuring the change in estimation accuracy on \joblr.  We use the smaller \sys in Table~\ref{table:joblr-error} as the \emph{Base} configuration, and ablate each component in isolation.  Table~\ref{table:ablation} presents the results.

In (A), using IBJS adapted for full joins\footnote{The fact table \textsf{title} is ordered at front and a large intermediate size of $10^6$ is used.} as a \emph{biased} sampler significantly decreases the learned estimator's accuracy.
The large increase in the median error implies a systematic distribution mismatch.
Overall, this design choice is the second most important. %

Rows in group (B) vary the column factorization granularity.  Using smaller bits results in more subcolumns and yields a small drop in accuracy.  Disabling factorization uses the most space and appears to perform the best.

Group (C) varies the size of the autoregressive model, by changing the dimension of the feedforward linear layers ($d_\text{ff}$) or the embeddings ($d_\text{emb}$).
An enlarged embedding proves markedly more useful than enlarged linear layers, likely because each token's captured semantics becomes more finetuned during optimization.

In group (D) we vary the correlation learned by \sys.
While all configurations above learn the distribution of all tables in a single model---capturing all possible correlations among them---here we build one model (same architecture as Base) per table. Queries that join across tables are estimated by combining individual models' estimates via independence.
Without modeling inter-table correlations, this variant yields the lowest accuracy.

Finally, group (E) ablates away the AR model altogether. We test \emph{uniform join samples} as a standalone estimator: it uses our sampler (\secref{sec:sampling}) to draw $10^4$ simple random samples (actual tuples in the database) from each query's join graph.
While the median error is reasonable, it is $10^4\times$ less accurate than an autoregressive model at tail as many queries have no sample hits.
The AR model is more statistically efficient than sampling, because it provides access to conditional probability distributions---these conditional contributions enable an efficient probabilistic inference procedure, i.e., progressive sampling, which cannot be used otherwise.

{\emph{Tuning guide.} Groups (B) and (C) show that \sys is not overly sensitive to hyperparameters. For new datasets, we recommend starting with the Base configuration and increasing sizing as much as possible up to a size budget. The recommended precedence is: factorization bits; $d_{\text{emb}}$; $d_{\text{ff}}$ and the number of layers.
  The number of training tuples can be set by early-stopping or a time budget; \secref{sec:efficiency}'s results suggest starting with a few to $10+$ million.
}

\begin{figure}[t]
  \centering
  \includegraphics[width=.98\columnwidth]{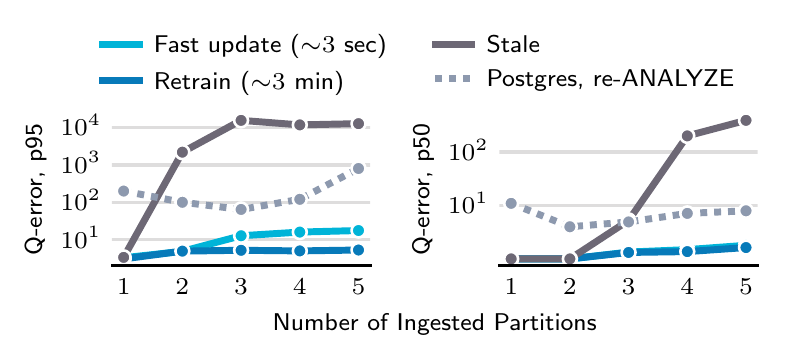}
\vspace{-.20in}
\caption{\small {Updating \sys, fast and slow. \jobl. {Errors (p95, p50) of each strategy are averaged from 10 runs.
          Postgres is also run as comparison, whose statistics are updated ($\bm{1\sim2}$ sec.) on each ingest.}}
\label{fig:data-updates}}
\vspace{-.35in}
\end{figure}

\subsection{Update Strategies}
\label{sec:updates}
\sys handles new data by either retraining, or taking additional gradient steps, i.e., incremental training.
To test both strategies, we simulate the practice of \emph{time-ordered partition appends}:
table {\sf title} is range-partitioned on a year column into 5 partitions. %
Each partition defines a distinct snapshot of the entire database and the full join, so running the same set of queries at different partition count yields 5 sets of true cardinalities.
We compare three update strategies, all of which are trained fully for 7M tuples after the first ingest: (1) \emph{stale}, trained once on the first snapshot and never updated, (2) \emph{fast update}, incrementally updated after each new ingest on 1\% of original samples (70K), and (3) \emph{retrain}, using 100\% of original samples (7M) after each ingest.
We also show the latency required to perform additional gradient steps.

Results are shown in {Figure~\ref{fig:data-updates}}.
Without update, the stale \sys significantly degrades in accuracy, which is expected as each partition adds a significant amount of new information.
A fast updated \sys recovers most of the accuracy, incurring a minimal overhead.
Even fully retraining only requires a few minutes and yields the highest accuracy.
Both the statistical efficiency (number of tuples needed vs. accuracy) and the physical efficiency of \sys contribute to these highly practical update strategies.

%% file: related.tex
\section{Related work}
\label{sec:related}

\vspace{6pt}
\noindent {\bf Unsupervised data-driven cardinality estimators.}
This family approximates the data distribution and dates back to System R's use of 1D histograms~\cite{systemr}.
The quality of the density model used has seen steady improvements throughout the years:

{\bf Classical methods.} Multidimensional histograms~\cite{poosala1997selectivity,gunopulos2005selectivity,muralikrishna1988equi,poosala_improved} are more precise than 1D histograms by capturing inter-column correlations.
Starting from early 2000s,
graphical models were proposed for {either single-table or join} cardinality estimation~\cite{getoor2001selectivity,tzoumas2011lightweight,deshpande2001independence}. These density models tradeoff precision for efficiency by assuming conditional or partial independence, and require expensive structure learning (finding the best model structure given a dataset).

{\bf Sum-product networks.} SPNs, a tree-structured density estimator, were proposed about 10 years ago~\cite{poon2011sum}.  Each leaf is a coarse histogram of a slice of an attribute, and each intermediate layer uses either $\times$ and $+$ to combine children information.  Due to their heuristics (e.g., inter-slice independence), SPNs have \emph{limited expressiveness}: there exists simple distributions that cannot be efficiently captured by SPNs of any depth~\cite{martens2014expressive}.
DeepDB~\cite{deepdb} is a recent cardinality estimator that uses SPNs. \sys is similar to DeepDB in the following aspects. \emph{(S1)} Both works use the formulation of learning the full outer join of multiple tables. \emph{(S2)} Our ``schema subsetting'' capability builds on their querying algorithms.

\sys differs from DeepDB in the following. \emph{(D1) Modern density model:} \sys's choice of a deep autoregressive model {is} a universal function approximator hence fundamentally more expressive.  Unlike SPNs, no independence assumption is made in the modeling.
\emph{(D2) Correlations learned:} \sys argues for capturing as much correlation as possible across tables, and proposes learning the full outer join of all tables of a schema.  DeepDB, due to limited expressiveness, learns multiple SPNs,
each on a table subset ($\sim 1$--3 tables) {chosen by correlation tests.}
Conditional independence is assumed across table subsets.
{\emph{(D3) Correct sampling:}}
\sys identifies the key requirement of sampling from the data distribution of joins in an unbiased fashion.  In contrast, DeepDB obtains join tuples either from full computation or IBJS which samples from a biased distribution.
Due to these differences, \sys outperforms DeepDB by up to $34\times$
in accuracy (\secref{sec:eval}).

{\bf Deep autoregressive models.}  A breakthrough in density estimation, deep AR models are the current state-of-the-art density models from the ML community~\cite{gpt2,made,transformer,resmade}.
They tractably learn complex, high-dimensional distributions in a neural net, capturing all possible correlations among attributes. %
Distinctively, AR models provide access to all conditional distributions among input attributes.
Naru~\cite{naru} is a single-table cardinality estimator that uses a deep AR model. By accessing conditional distributions, Naru proposes efficient algorithms to integrate over an AR model, thereby producing selectivity estimates.  \sys builds on single-table Naru and overcomes the unique challenges (\secref{sec:overview}) to support joins. %

\vspace{4pt}
\noindent {\bf Supervised query-driven cardinality estimators.}
Leveraging past or collected queries to improve estimates dates back to LEO~\cite{leo}.
Interest in this approach has seen a resurgence partly due to an abundance of query logs~\cite{msrlearnedcard} or better function approximators (neural networks)~\cite{kipf2018learned,e2e_cost} that map featurized queries to predicted cardinalities.
Hybrid methods that leverage query feedback to improve density modeling have also been explored, e.g., KDE~\cite{selectivity-kde,kiefer2017estimating} and mixture of uniforms~\cite{quicksel}.
Supervised estimators can easily leverage query feedback, handle complex predicates (e.g., UDFs), and are usually more lightweight~\cite{msr-lightweight}.
\sys has demonstrated superior estimation accuracy to representatives in this family, while being fundamentally more robust since it is not affected by out-of-distribution queries.
Complex predicates can also be handled by executing on tuples sampled from \sys's learned distribution.

\vspace{4pt}
\noindent {\bf Join sampling.} Extensive research has studied join sampling, a fundamental problem in databases.
\sys leverages a state-of-the-art join sampler to obtain training tuples representative of a join.
\sys adopts the linear-time Exact Weight algorithm from Zhao \etal~\cite{zhao_sampling}, which is among the top-performing samplers they study.
This algorithm provides uniform and independent samples, just as \sys requires.
\sys may further leverage their extensions to support cyclic join schemas.
While IBJS~\cite{ibjs_cite} and Wander Join~\cite{WanderJoin} provide unbiased estimators for counts and aggregates, they do not provide uniform samples of a join and thus are unsuitable for collecting training data. Lastly, we show that it is advantageous to layer a modern density model on join samples.

\vspace{4pt}
\noindent {\bf Learned database components.}  A great deal of work has recently applied either classical ML or modern deep learning to various database components, e.g., indexing~\cite{kraska2018case}, data layout~\cite{qdtree}, and query optimization~\cite{dq,skinnerdb_sigmod,neo}.  \sys can be seen as a versatile \emph{core} that can benefit any query engine, learned or not learned.  Being able to model inter-table and inter-column correlations without any independence assumptions,
\sys's use may go beyond query optimization to other tasks that require an \emph{understanding of tables and attributes} (e.g., data imputation~\cite{wu2020attention} or indexing~\cite{wu2019designing}). %

%% file: conclusion.tex
\section{Conclusion}
\label{sec:conclusion}

\sys is built on a simple idea: learn the correlations across all tables in a database without making any independence assumptions.
\sys applies established techniques from join sampling and deep self-supervised learning to cardinality estimation, a fundamental problem in query optimization.
It learns from data---just like classical data-driven estimators---but captures all possible inter-table correlations in a probabilistic model: $p_\theta(\text{all tables})$.
To our knowledge, \sys is the first cardinality estimator to achieve assumption-free probabilistic modeling of more than a dozen tables.
\sys achieves state-of-the-art accuracy for join cardinality estimation
(4--34$\times$ better than prior methods)
using a single per-schema model that is both compact and efficient to learn.